\documentstyle[12pt,aaspp4]{article}

\lefthead{Cowley et al.}
\righthead{Supersoft X-ray Binaries}

\begin{document}

\title{SIX SUPERSOFT X-RAY BINARIES: SYSTEM PARAMETERS AND TWIN-JET OUTFLOWS}

\author{A.P. Cowley\altaffilmark{1} and P.C. Schmidtke\altaffilmark{1} }
\affil{Department of Physics \& Astronomy, Arizona State University,
Tempe, AZ, 85287-1504} 

\author{David Crampton\altaffilmark{1} and J.B.
Hutchings\altaffilmark{1}} \affil{Dominion Astrophysical Observatory,
National Research Council of Canada,\\ Victoria, B.C. V8X 4M6, Canada} 

\altaffiltext{1} {Visiting Astronomers, Cerro Tololo Inter-American
Observatory, National Optical Astronomy Observatories, which is operated
by the Association of Universities for Research in Astronomy, Inc., under
contract with the National Science Foundation}

\begin{abstract}

A comparison is made between the properties of CAL 83, CAL 87, RX
J0513.9$-$6951, 1E 0035.4$-$7230 (SMC 13), RX J0019.8+2156, and RX
J0925.7$-$4758, all supersoft X-ray binaries.  Spectra with the same
resolution and wavelength coverage of these systems are compared and
contrasted.  Some new photometry is also presented.  The equivalent widths
of the principal emission lines of H and He II differ by more than an
order of magnitude among these sources, although those of the highest
ionization lines (e.g. O VI) are very similar.  In individual systems, the
velocity curves derived from various ions often differ in phasing and
amplitude, but those whose phasing is consistent with the light curves
(implying the lines are formed near the compact star) give masses of
$\sim$1.2M$_{\odot}$ and $\sim$0.5M$_{\odot}$ for the degenerate and
mass-losing stars, respectively.  This finding is in conflict with
currently prevailing theoretical models for supersoft binaries.  The three
highest luminosity sources show evidence of ``jet" outflows, with
velocities of $\sim$1--4$\times$10$^3$ km s$^{-1}$.  In CAL 83 the shape
of the He II 4686\AA\ profile continues to show evidence that these jets
may precess with a period of $\sim$69 days. 

\end{abstract}


\section{INTRODUCTION}

The close-binary ``supersoft sources" (SSS) are now recognized as a
distinct class of very luminous (L$_{bol}\geq10^{38}$ erg s$^{-1}$) X-ray
sources characterized by extremely soft X-ray spectra with little or no
radiation above $\sim$0.5 keV (e.g. Tr\"umper et al.\ 1991).  Several
reviews of the observational properties of these sources have recently
been published (e.g. Hasinger 1996, Greiner 1996, Kahabka \& van den
Heuvel 1997).  All SSS appear to have high mass-accretion rates and
exhibit long-term X-ray and optical variability which are thought to
reflect variations in the rate of mass transfer.  In addition, some SSS
show evidence of collimated outflows or ``jets" (Crampton et al.\ 1996,
hereafter CHC96; Southwell, Livio \& Pringle 1997).  Van den Heuvel et
al.\ (1992) suggested that the X-ray properties of SSS are best explained
by a model involving steady nuclear burning on the surface of a white
dwarf accreting material at the Eddington rate.  Many observations appear
to support this model (Greiner 1996), although alternative interpretations
(e.g. Kylafis \& Xilouris 1993) have not yet been ruled out. 

During a 1996 November CTIO observing run we obtained spectra and some
photometry of six close-binary supersoft sources.  One of these lies in
the Small Magellanic Cloud, 1E 0035.4$-$7230 (hereafter SMC 13).  CAL 83,
CAL 87, and RX J0513.9$-$6951 (hereafter RX J0513) are all members of the
Large Magellanic Cloud, while RX J0019.8+2156 (hereafter RX J0019) and RX
J0925.7$-$4758 (hereafter RX J0925) are galactic systems.  Since these
objects were observed with the same spectrographic configuration,
intercomparison of their spectra is very straightforward.  In addition, we
present previously unpublished photometry for CAL 83 and a few
observations of RX J0513 and RX J0925.  New data for CAL 87 is being
published in a separate paper (Hutchings et al.\ 1998).  Long-term
spectroscopic and photometric monitoring of these sources is important
since the SSS exhibit significant variations over timescales of months and
years. 

A summary of the properties of these six supersoft binaries is given in
Table 1, where they are listed in order of decreasing orbital period.  The
bolometric luminosity (L$_{bol}$) listed is the average value given in
Greiner's catalog (1996); it depends strongly on the assumed model,
adopted distance, and amount of absorption assumed.

\section{OBSERVATIONS AND MEASUREMENTS}

\subsection{Spectroscopic Data}

Spectra of the SSS were obtained with the CTIO 4-m telescope during five
nights in 1996 November with the KPGL1 grating and Loral 3K detector.  The
spectra cover the wavelength range 3700 -- 6700\AA\ and have a resolution
of $\sim$1.0\AA\ per pixel.  With a 1\farcs5 slit, corresponding to three
pixels, the spectral resolution is $\sim$3\AA.  Due to the range in
magnitudes of these systems, quite different total exposure times and
phase coverage were achieved, but the spectrographic set-up and hence
resolution, etc., was the same.  The heliocentric Julian dates (HJD) of
mid-exposure, exposure times, and phases (where known) for four of the
sources are listed in Table 2.  Complete phase coverage was obtained for
the other two, SMC 13 and CAL 87, but details of these spectra are given
elsewhere (Crampton et al.\ 1997, Hutchings et al.\ 1998). 

One-dimensional spectra were extracted and processed following standard
{\sc IRAF} techniques.  He-Ne lamp spectra were taken before and after
each stellar exposure to calibrate the wavelengths which are established
to $\sim$0.1 pixel ($\pm$6 km s$^{-1}$) or better.  The
wavelength-calibrated spectra have a peak signal-to-noise (S/N) of
$\sim$12 per pixel. 

\subsection{Photometric Data}

Multicolor CCD photometry of the southern sources was obtained with the
0.9-m telescope at CTIO.  The images were reduced using DAOPHOT (Stetson
1987) and calibrated using observations of Landolt (1992) standard stars.
To improve photometric accuracy, differential measurements were made
relative to comparison stars within the field of view (see Schmidtke
1988).  Table 3 presents all of our previously unpublished photometry for
the southern SSS, taken during four observing runs between 1993 December
and 1996 November. 

$BV$ measurements of the galactic source RX J0019 were obtained from 1995
June 23 through 1996 January 14 using a 0.75-m automated photometric
telescope (APT) at Fairborn Observatory, then located on Mt.\ Hopkins,
Arizona.  Details regarding a similar telescope and single-channel
photometer are given by Pyper et al.\ (1993).  The source was monitored up
to three times per night, following a prescribed sequence that included
observations of the variable star (RX J0019), comparison star (SAO 73882,
$V=8.809$, $B-V=+0.482$), check star (SAO 73903, $V=9.164$, $B-V=+0.497$),
and a nearby sky patch.  Differential photometry of RX J0019 was
calculated using only those sequences in which the standard error of the
mean for both variable and comparison star was $<$0.02 mag in each filter.
Two observations that met these internal consistency checks displayed
large random errors and were removed from the data set, yielding 115 $BV$
measurements on 66 nights.  As a result of the 0.02-mag constraint,
differentials taken during times of rapid flickering are ignored.  APT
data are most useful in assessing the general behavior of the RX J0019
light curve. 

\section{DESCRIPTION OF INDIVIDUAL OBJECTS}

In this section we describe our spectra of individual objects and compare
them with previous observations.  To facilitate intercomparison, all
spectra were normalized using identical methods.  The resulting spectra of
the six SSS are shown in Figures 1a and 1b, in order of decreasing
emission-line intensity.  For display purposes, the spectra were smoothed
with a 1.5\AA-width gaussian, and the wavelength scales of the galactic
sources were shifted to match the Magellanic Cloud sources.  Several of
the prominent features are marked, as well as some of the strongest
diffuse interstellar absorption bands (DIB).  The SSS all vary both with
phase and over longer time intervals, so these spectra are just snapshots
of their spectral appearance.  Furthermore, the galactic source RX J0925
is heavily reddened so the S/N in the short-wavelength region of its
spectrum is low. 

A discussion of each source is given in the following sections.  For
simplicity, they are arranged below in order of right ascension: RX
J0019.8$+$2156 (galactic), SMC 13 $=$ 1E 0035.4$-$7230 $=$ RX
J0037.3$-$7214 (SMC), RX J0513.9$-$6951 (LMC), CAL 83 $=$ RX
J0543.7$-$6822 (LMC), CAL 87 $=$ RX J0546.9$-$7108 (LMC), and RX
J0925.7$-$4758 (galactic). 

\subsection{RX J0019.8$+$2156}

RX J0019 is one of the two galactic supersoft sources which we observed. 
Because its distance is not well known, the values of its absolute
magnitude and some other quantities given in Table 1 are not well
established.  The system lies between 1 and 2 kpc.  Following Greiner 
(1996), we have used a distance of 2 kpc to compute values in the table.

Two spectra of RX J0019 were obtained, and neither shows any evidence for
spatially extended emission lines in our 2-dimensional images, even though
the system is close enough so that it might be possible to detect a modest
surrounding nebula if one were present.  The spectra were taken at
spectroscopic phases 0.39 and 0.47, where phase zero is defined as maximum
positive velocity (i.e. standard spectroscopic notation).  To compute
these phases we adopted the photometric ephemeris of Will \& Barwig (1996)
(P $=$ 0.6604721 days, T$_0$(phot) $=$ HJD 2448887.5091) and added 0.25P
to that value.  Beuermann et al.\ (1995) have shown that this is
approximately correct for this system.  We have added our two exposures to
give the mean spectrum shown in Figure 1.  Note that the spectrum closely
resembles that of CAL 83, except that the 4640--50\AA\ blend is much
weaker.  Beuermann et al.\ (1995) have shown that the systemic velocity is
$-59$ km s$^{-1}$ and the semi-amplitude is K$=$67 km s$^{-1}$.
Measurement of our strong emission lines gives a velocity of $-94$ km
s$^{-1}$, indicating the source was observed near quadrature, as the
computed spectroscopic phases also indicate. 

Satellite lines are visible on both sides of the strongest emission lines
(see Figure 1).  He II 4686\AA\ shows both positive and negative emission
components of about equal strength.  The H Balmer lines and He II
Pickering lines show the shortward displaced component in absorption.  In
H there is a sharp absorption while in the He II lines there is a much
weaker absorption which appears as a steep edge to the emission, but it is
not seen as a separate line.  The longward component is in emission at the
Balmer lines but not seen at the Pickering lines.  The O VI lines do not
show either displaced component.  All of these high-velocity lines, whether
emission or absorption, show similar offsets from the central emission
component.  From the six strongest lines, the average shortward velocity
relative to the main line is $-$690 km s$^{-1}$ and the longward velocity
is $\sim +$712 km s$^{-1}$, which are the same to within the $\pm35$ km
s$^{-1}$ measurement error.  These observations suggest the presence of a
double-sided (bipolar) jet, as has previously been found in RX J0513
(CHC98, Southwell et al.\ 1996).  The difference in the strengths of the
absorption component between various ions could indicate that the
temperature of the jet falls with distance from the star, so that in the
outer regions of the jet the lowest ionization lines are seen in
absorption.  Modeling of the light curve for RX J0019 gives an orbital
inclination of $i=56^{\circ}$ (Schandl et al.\ 1996).  By comparison with
RX J0513 with its low orbital inclination and high-velocity jets, one
might expect a system with a higher inclination to show lower-velocity
jets since they are not pointed towards the observer. 

Two high dispersion spectra of RX J0019 were obtained at the MMT by Mark
Wagner on 1995 October 12 and 1996 August 4.  The 1995 spectrum shows no
evidence of displaced lines at He II 4686\AA, but at H$\alpha$ there is a
weak, positively displaced emission component at $\sim+800$ km s$^{-1}$. 
On the short-wavelength side of the line there is a moderately broad
absorption centered at $\sim-800$ km sec$^{-1}$.  The 1996 August spectrum
shows stronger jet lines, with both positively and negatively displaced
emissions at He II 4686\AA\ \underbar{and} He II 5411\AA.  H$\alpha$,
H$\beta$, H$\gamma$, and H$\delta$ all show an emission at $\sim+830$ km
s$^{-1}$ and evidence that there is a similar negatively displaced
emission, but it which is strongly divided by a fairly narrow absorption
at $\sim-450$ km sec$^{-1}$.  Thus, the appearance of the jet lines changed
considerably in about ten months, with the displaced features being more
prominent in 1996 August than in 1995 October.  By 1996 November these
displaced lines had become even stronger and could easily be seen on the
moderate-resolution spectra taken at CTIO. 

A series of spectra taken during two consecutive nights in 1995 September
at Lowell Observatory show no evidence of jet lines, but the spectra did
not include the H$\alpha$ region.  This is in agreement with the
high-dispersion 1995 October MMT spectrum which showed no displaced lines
at He II 4686\AA.  Thus, over a period of about a year either the strength
of the jets changed, or at times the velocity of these jet-like features
may change, perhaps due to precession, causing them to merge with the
central emission. 

Greiner \& Wenzel (1995) demonstrated that the long-term ($\sim$100 years)
light curve of RX J0019 displays irregular variability with an amplitude
of $\sim1$ mag on timescales of several years and smaller variations in a
few weeks.  We do not know whether the spectral changes are correlated with
the long-term photometric changes. 

We have obtained some photometric data covering the period 1995 June to
1996 January with the APT telescope on Mt.\ Hopkins.  This $V$-band
photometry confirms that RX J0019 is still variable on timescales of
months.  We have identified three levels of mean intensity in the APT
data, with the source being in a `low' state on November 10--30, a
`medium' state during October 12--28, and a `high' state for all remaining
observations.  Because the difference between high and medium states is
small, further monitoring is needed to confirm whether there is a true
distinction.  Figure 2 shows the RX J0019 light and color curves, folded
on the ephemeris of Will \& Barwig (1996).  Data for the three states are
represented by different symbols, with $V$ observations from the low and
medium states adjusted by 0.220 and 0.103 mag, respectively, to bring them
up to the high-state scale.  The shape of the $V$ light curve does not
vary from state to state.  It is dominated by a very broad, asymmetrical
primary eclipse, having more scatter during ingress (near phases 0.7--0.8)
than egress.  This behavior strongly resembles that of CAL 87 (see
Schmidtke et al.\ 1993) although the depth of primary minimum is
significantly less ($\sim$0.5 mag for RX J0019 vs.\ $\sim$2.0 mag for CAL
87).  Our high-state magnitude range ($V=12.4-12.9$) closely matches the
photometry of Matsumoto (1996), but differs from that given by Beuermann
et al.\ (1995) whose 1992 September 15--22 data are $\sim$0.25 mag
brighter than our high-state values at all phases. 

The bottom panel of Fig.\ 2 shows the ($B-V$)-color curve for RX J0019.
Unlike the $V$ data, no shift has been applied since the mean color
($B-V=+0.008$) does not vary significantly between states.  However, the
curve has an unusual orbital modulation, with the source being {\it
bluest} during both primary and secondary minima.  Because the pattern is
present in data from all three states, it is not an artifact of combining
low, medium, and high-state observations.  Fitting a sinusoid to the $B-V$
measurements yields a peak-to-peak amplitude of 0.024 mag, which is much
larger than the internal consistancy of the data set.  Differential
photometry between the check and comparison stars is constant for both $V$
and $B-V$, with a 1-$\sigma$ dispersion of only 0.005 mag in each
parameter.  Hence, the observed color variations for RX J0019 appear to be
real and not caused by variations in the comparison star. 

Our APT data and extensive photometry obtained in 1995 September at Lowell
Observatory (Schmidtke et al., in preparation) also show short-term
changes in the light curve from cycle to cycle and night to night.  Similar
behavior has been reported by Meyer-Hofmeister et al.\ (1998) in
observations taken between 1996 October and December. 

The mean color of RX J0019 ($B-V\sim0.01$) indicates that the system is
slightly reddened.  Using the same $B-V$ color as for RX J0513
($B-V=-0.11$) implies absorption of A$_V\sim0.36$ for RX J0019.  From our
out-of-eclipse magnitude of $V\sim$12.4 and adopted a distance of 2 kpc
(Greiner 1996), one obtains M$_V\sim+0.6$, with considerable uncertainty. 

\subsection{1E 0035.4$-$7230 = SMC 13}
  
Of the observed SSS, SMC 13 is the faintest, has the shortest period, and
shows the weakest H and He II emission lines, although the O VI lines
(3811, 5290\AA) are comparable in strength to other SSS.  Presumably the
weakness of H and He II is due to its necessarily small accretion disk in
this short-period (0.1719 days; Schmidtke et al.\ 1996) system.  The O VI
lines which are formed in the innermost disk do not appear to be affected.
Crampton et al.\ (1997) carried out a spectroscopic analysis based on 17
spectra obtained in 1996.  These spectra have been averaged and are shown
in Figure 1.  As in CAL 87, the emission lines are noticeably broader than
for the other SSS.  SMC 13 and CAL 87 are the two systems with the highest
orbital inclinations (see Table 1), so the breadth of the lines must
result from viewing the rotating disk nearly edge-on. 

While the emission-line spectrum of SMC 13 is similar to other supersoft
sources, there is also a weak, broad Balmer absorption when the spectra
are binned by phase.  He II emission yields a well-defined velocity curve
whose phasing suggests the lines are formed near the compact star and
reveal its orbital motion.  Modeling of the light curve by
Meyer-Hofmeister et al.\ (1997) indicates that the system is marginally
eclipsing, with $i = 75^{\circ}$.  Using the He II emission velocity
amplitude to derive the mass function, we determine masses of
0.4M$_{\odot}$ for the donor and 1.3M$_{\odot}$ for the compact star,
quite different from those expected in the scenario of van den Heuvel et
al.\ (1992). 
 
The puzzle in this system is the Balmer absorption.  It is present in all
phase bins.  The lines show a high-velocity amplitude, with K$\sim400$ km
s$^{-1}$.  Its phasing is not well defined although compatible with being
due to orbital motion of the compact star.  However, the large amplitude
implies improbably high masses for both components.  The different
amplitudes of the absorption and emission lines suggest they originate at
different distances from the center of mass.  Since the absorption be must
seen against a continuum source, it is hard to imagine a geometry that
produces this, unless the observed absorption-line velocity curve is not
related to any orbital motion.  The high absorption-line velocities cannot
be due to gas outflow since the mean velocity would be expected to be more
negative than that from the emission lines, which is it not.  Thus, the
`best' radial velocity curve in SMC 13 is based on the emission lines.  We
do not understand the origin or interpretation of the absorption lines. 

\subsection{RX J0513.9$-$6951}

The spectrum of RX J0513 has been described by Pakull et al.\ (1993),
Cowley et al.\ (1993), CHC96, and Southwell et al.\ (1996).  RX J0513
exhibits the strongest emission lines of all the SSS and also shows weak,
highly shifted emission features which arise from collimated outflows or
``jets" (CHC96, Southwell et al.\ 1996).  The 1996 spectra are similar to
those taken in 1994 (CHC96), although the detailed shapes of the lines,
including those of the jet lines, show changes. 

An intercomparison of our 1993, 1994 and 1996 spectra in the spectral
region 4600--5000\AA\ is shown in Figure 3.  The long-wavelength wings of
the H$\beta$ and He II 4686\AA\ are stronger in 1994 and 1996 than in 1993
(also see Fig.\ 7 of CHC96) as is the complex of C III, N III, and N V
lines shortward of 4686\AA.  In 1996 the highly displaced ``jet" lines are
much weaker and may have lower velocities than in 1994.  The
violet-shifted He II (4686\AA) emission line is the best defined of the
jet features.  In order to compute the outflow velocities for these lines,
we have adopted a systemic velocity for RX J0513 of $+280$ km s$^{-1}$ and
determined the displacement from a central line with that velocity.  In
the 1996 spectra the negatively displaced He II emission has a wavelength
of 4634.9\AA\ ($-3529$ km s$^{-1}$) compared with 4628.0\AA\ ($-$3971 km
s$^{-1}$) in 1994.  In the 1996 spectra there is a relatively strong
``P-Cygni" absorption feature at 4809.8\AA, compared to its being mostly
in emission in 1994.  Since only hydrogen lines show absorption
components, we identify this displaced feature with H$\beta$ giving, it an
outflow velocity of $-3458$ km s$^{-1}$, similar to the He II
high-velocity emission line.  In the 1996 spectra, there is a sharp
emission feature at 4935.5\AA.  If it is the red-shifted component of
H$\beta$, it has an outflow velocity of $+$4296 km s$^{-1}$, but the width
of the feature in 1993 and 1994 suggests they could be blends.  Hence the
1996 feature could be entirely due to an unidentified emission line which
contributes to the blend.  The corresponding positive component of
4686\AA, which should be near 4760\AA, is not present in our 1996 spectra.

In Figure 4 the 1996 H$\alpha$ and H$\beta$ line profiles are compared
with the 1994 H$\beta$ line.  These spectra are displayed on a velocity
scale to show that features with similar velocities are present at both
H$\alpha$ and H$\beta$.  In the 1996 spectra there is a highly displaced
absorption line at 6496\AA\ ($-$3332 km s$^{-1}$ with respect to the
systemic wavelength of H$\alpha$) which has a shape and velocity similar
to the one near H$\beta$.  Thus, the negatively displaced He II 4686\AA\
emission line and the absorption features shortward of the H lines have
comparable velocities.  There appears to be an emission line at
$\sim$6656\AA\ (outflow velocity of $+$3976 km s$^{-1}$ if it is the
red-shifted emission component of H$\alpha$).  However, this feature is so
close to the end of the spectrum, its wavelength is not reliably
determined.  The lack of a corresponding He II component makes it
uncertain if the feature is the positively displaced jet line. 

In an effort to look more closely at the asymmetry of the 1996 emissions
at H$\alpha$ and H$\beta$, we have reflected the long-wavelength side of
the emission profile onto the short-wavelength side and then subtracted
the blueward profile.  The difference between the two sides show evidence
of absorption at velocities of $\sim-500$ to $-1700$ km s$^{-1}$ and 
$\sim-2200$ to $-4000$ km s$^{-1}$.  This, of course, assumes a
symmetrical emission profile which is defined by the redward side.  In
1994 the hydrogen emission was more symmetrical.  Doing a similar
reflection and subtraction to the 1994 spectra shows the high velocity
absorption was weakly present.  Thus apparently there are both low and
high velocity absorption components at the hydrogen lines.  This behavior
may be analogous to stellar winds where the densities and velocities
change as the material is accelerated outward. 

Thus, comparison of our 1996 and 1994 spectra show that the principal
emission lines (He II and H) have similar strengths although their
profiles have changed.  The O VI lines have comparable strengths in both
years, within our measurement errors, while the complex of lines near
4650\AA\ (C III, N III, C IV) is considerably stronger in the 1996
spectra.  The negatively displaced ``P-Cygni" H absorption lines are
stronger in 1996, although they were weakly present in 1994, while the
positively displaced ``jet" emission lines are weaker in 1996 than in
1994.  The change in the strength of the displaced absorption component
suggests that our line-of-sight in 1996 may have been more along the
approaching jet causing the features to be seen in absorption. 

In this source the line of sight is near the polar axis
($i\sim15^{\circ}$), and thus we may be looking along along the jets.  The
fact that the shortward-shifted component is seen in absorption suggests
that either the disk or the jet emission itself lies behind the absorber;
both require our line-of-sight to be almost parallel to the jet.  We would
not expect to see high-velocity absorption in any other situation, so that
this is a further indication that the inclination of the system is very
low.  The absorption profile, as in a stellar wind, shows a terminal
velocity close to that seen in the receeding jet emission.  This means we
may be looking through an accelerating region of the jet, going from
$\sim$90\% to its final value.  The variation of the profile and depth of
the absorption over timescales of a year indicate that the absorbing
column or the velocity profile in the outflow change, or that our
line-of-sight is not always directly along the jet (perhaps due to
precession). 

Alcock et al.\ (1996) and Reinsch et al.\ (1996) have published extensive
photometry of RX J0513 which shows that the system exhibits recurring low
states, fading irregularly by $\sim$1 mag every few months.  Our 1992,
1994 and 1996 spectra all appear to have been taken when it was in its
normal ``high" state.  Our photometry (Crampton et al.\ 1996) confirms
that the source was in a high state when our 1994 spectra were obtained. 
One 1995 November image also shows the source to have been in its bright
state, with $V=16.8$.  This is in good agreement with overlapping data of
Alcock et al.\ (1996).  We do not have any 1996 photometry for RX J0513,
but the spectral appearance suggests the source was bright. 

By contrast, during our 1993 observations RX J0513 was in a low state
(Reinsch et al.\ 1996), and the principal emission lines of H and He II
were weaker.  Reinsch et al.\ report similar observations and explain both
the emission-line and photometric behaviors with a simple model in which
decreasing mass transfer causes the photosphere of the accreting white
dwarf to shrink, thus reducing the illumination of the disk which is
responsible for most of the optical emission. 

Using the extensive MACHO photometric data, Alcock et al.\ (1996) detected
a small orbital variation with a semi-amplitude of $\sim$0.02 mag and a
period P$=$0\fd76278.  (As an aside, Alcock et al.\ give T$_0$ as time of
\underbar{maximum} light rather than the usual notation as time of minimum
light.)  Their value of the period is consistent, within the errors of
both determinations, with the spectroscopic period found by Crampton et
al.\ (CHC96) of 0\fd75952 and later confirmed by Southwell et al.\ (1996).
Because of the very small range of the light variation, T$_0$ is poorly
determined.  Therefore, to compute phases listed in Table 2 we have used
the Alcock et al.\ value for the period but the epoch of maximum radial
velocity from the CHC96 spectroscopic study (T$_0$(max vel)$=$HJD
2449332.63).  Using these phases, our 1996 November spectra are expected to
have velocities slightly above the systemic value, and our measurements
show that they do. 

The very small amplitudes of both the orbital velocity and light
variations indicate that RX J0513 is viewed nearly pole-on.  If the
emission lines (and most of the light) originate in the accretion disk,
then for the simplest case minimum light should occur at spectroscopic
phase 0.25.  However, minimum light (as defined by T$_0+0.5$P, using the
Alcock et al.\ T$_0$) occurs at spectroscopic phase $0.05\pm0.03$.  In
such a low inclination system, minimum light is probably not a good
indicator of when the stars are at conjunction, but rather when some disk
structure or stream occults some of the light from the inner disk. 
However, if the light curve really did reveal the relative orientation of
the stars, then the observed velocities would all be due to non-orbital
motions. 

To summarize, RX J0513 is brightest of the supersoft binaries we observed.
This may imply that it is undergoing the highest rate of mass transfer. 
It shows the most pronounced high-velocity jets of all the SSS we have
observed, but of course its low orbital inclination makes detection of
these more favorable than in high-inclination systems. 

\subsection{CAL 83 $=$ RX J0543.7$-$6822} 

The spectrum of CAL 83 has been described by Pakull, Ilovaisky, \&
Chevalier (1985), Crampton et al.\ (1987), and Smale et al.\ (1988).  
The latter authors determined the orbital period to be 1.044 days from their
extensive photometric observations.  Based on spectroscopic observations
obtained between 1982 and 1987, the period was later refined to be 1.0475
days (Cowley et al.\ 1991) with the epoch of maximum positive velocity
being T$_0$(max vel) $=$ HJD 2445727.70.  All three of our 1996 spectra
were taken near phase $\Phi_{spec} =$ 0.5, based on this ephemeris.  Our
velocities fit the velocity curve determined by Crampton et al.\ (1987)
quite well.  The mean 1996 spectrum is very similar to that from 1994,
although the He II and Balmer lines have increased somewhat in strength. 

Crampton et al.\ (1987) reported long-term changes in the wings of the He
II and H emission lines.  They suggested material may be ejected in
bipolar flows from a precessing disk which has a possible 69-day period. 
Our 1996 spectra (see Figure 1) show the red wings of the emission lines
were enhanced.  All of our 1994 and 1996 spectra are still consistent with
the proposed 69-day period.  In addition, a spectrum shown by Charles et
al.\ (1997) which was taken on 1997 January 5 also has a pronounced
high-velocity red extension to the emission lines.  This appearance also
fits the same 69-day period.  However, this long period is based on 11
epochs widely spaced over 15 years, so it is still rather poorly
constrained. 

Recently, Alcock et al.\ (1997) reported photometric observations
contemporaneous with an X-ray turn-off observed by $ROSAT$ in 1996 April 
(Kahabka 1996).  The optical decline occurred $\sim$10--15 days before the
X-ray minimum.  This behavior is expected if the optical luminosity arises
mostly from the accretion disk.  As the rate of mass transfer decreases,
the optical luminosity decreases.  The steady nuclear burning ceases which
causes the hot surface layers of a white dwarf to cool and contract
resulting in the X-rays turning off.  They conclude that to explain their
observations with this model, the mass of the white dwarf must be quite
high, M$_{WD}>1.3$M$_{\odot}$. 

Based on the velocity variation of He II emission, Crampton et al.\ (1987)
showed that for any value of the compact-star mass, the unseen secondary
star is less massive.  Based on both the light curve and emission-line
widths, the orbital inclination of CAL 83 must lie in the range
$i=20-30^{\circ}$.  If the compact star is a 1.3M$_{\odot}$ white dwarf
(Alcock et al.), then the secondary star must have a mass of
$\sim$0.5M$_{\odot}$, again considerably smaller than the
$\sim$2M$_{\odot}$ star expected by the van den Heuvel et al.\ (1992)
model. 

Table 3 presents unpublished photometry of CAL 83, primarily from 1995
November and 1996 November, although one point was measured in 1993
December.  Combining these data with previously published photometry, it
is apparent that the mean light level of CAL 83 varies considerably from
season to season, probably similar to the behavior of RX J0513 which is
so well documented.  Table 4 summaries the average magnitudes in different
epochs going back to 1980.  However, there are insufficient CAL 83 data to
determine if the high and low states vary regularly (perhaps related to
the 69-d spectroscopic variations) or irregularly as in RX J0513. 

Pakull \& Motch (1989), Bland-Hawthorn (1995), and Remillard, Rappaport,
\& Macri (1995) have published [O III] images of the nebula around CAL 83
which appears to be ionized by the soft X-ray flux from the central
source.  Rappaport et al.\ (1994) presented detailed calculations of the
ionization and temperature structure expected in such nebulae.  Remillard,
Rappaport, \& Macri (1995) gave an analysis of spectra similar to our own
as well as their [O III] and H$\alpha$ images, making detailed comparison
with their theoretical models.  Our spectra, taken with a 1\farcs5
$\times$ 90\arcsec~ slit centered on CAL 83, show emission extending over
$\sim$75\arcsec~ with enhancements at $\pm13\arcsec$ from the central star
in all lines except He II 4686\AA.  A portion of one of our
two-dimensional spectra is displayed in Figure 5.  The intensity of He II
4686\AA\ declines radially outward from the star.  Extended emission is
also visible at [O II] 3727\AA, H$\gamma$, H$\beta$, [O III] 4959\AA\ +
5007\AA, and H$\alpha$ (only He II 4646\AA, H$\beta$, and [O III] are
displayed in Fig. 5).  H$\alpha$ is weaker than expected, probably because
it was partially removed by our background subtraction procedure (based on
flux near the ends of the slit) since it has a larger spatial extension.
Nevertheless, the spatial variation over our $90\arcsec$ slit is extremely
small, much less than for the other lines.  In the figure note the very
different distributions of intensity along the slit for He II 4686\AA\
compared to H$\beta$ and the [O III] lines. 

\subsection{CAL 87 $=$ RX J0546.9$-$7108}
 
CAL 87 is unique in showing deep eclipses which, in principle, provide
important constraints on the system parameters.  Ironically, there are
many uncertainties in the interpretation of the velocities of various
observed lines.  This is caused partially by the high inclination since at
this orientation one detects the complex motions in the accretion disk.
 
The light-curve analysis carried out by Schandl et al.\ (1996), using a
model with an azimuthally irregular disk, yields an orbital inclination of
78$^{\circ}$.  Regardless of the light-curve model used, the inclination
must be larger than $\sim65^{\circ}$ to cause eclipses of any kind.  When
orbital inclinations are this high, the masses derived through the mass
function change little with inclination. 
 
In our original study of CAL 87, He II 4686\AA\ was measured from a series
of spectra observed in 1988 December (Cowley et al.\ 1990).  The resulting
radial velocity curve showed the He II emission line to have a
semi-amplitude of K$=$40 km s$^{-1}$ which was properly phased with
respect to the light curve to suggest it might be formed near the compact
star.  Using this low amplitude to compute the mass function resulted in
large masses  for the compact star ($>4$M$_{\odot}$) for any donor mass
$>0.4$M$_{\odot}$.  A few spectra obtained in 1993 showed quite different
phasing and amplitudes, suggesting that the system undergoes significant
changes, but there were not enough data to quantify the changes or to
redetermine the masses. 
 
In 1996 we obtained a major new dataset (Hutchings et al.\ 1998) from
which we have derived more extensive spectroscopic information, but these
data still do not finally resolve the question of the masses.  The
spectrum shown in Figure 1 is a simple average of the 25 spectra taken in
1996, scaled by their average fluxes.  The individual spectra show strong
Balmer absorption lines which vary with phase, as well as He II and O VI
emission and Ca II absorption.  All lines have different amplitudes and
phasing.  The phasing of the velocity curves may result from a combination
of superimposed absorption on the emission lines distorting their
profiles, azimuthally varying obscuration in the disk, and non-orbital
motions in the system.  The Balmer absorptions are the only velocities
that are phased a quarter cycle away from the eclipse, so that they might
reflect the orbital motion of the compact star.  Their semi-amplitude is K
$=$ 73 km s$^{-1}$ which, if orbital, implies masses similar to those
found for SMC 13, as discussed above.  A massive white dwarf
(1.3M$_{\odot}$) would have a donor star with a mass of only $\sim$0.4
M$_{\odot}$ in this system.  Again, this differs from the `standard' model
in which the nondegenerate star is expected to be in the range
1--2M$_{\odot}$.  The O VI and He II emission lines have lower K values
(near $\sim$30 km s$^{-1}$), but the maximum of the velocity curve occurs
about 0.14P too early, suggesting contamination with non-orbital motions.
Since the phasing is rather poorly constrained, we can consider what
masses these rather high excitation lines which must be formed near the
compact star would give.  They imply donor star masses of about
0.4M$_{\odot}$ and a massive compact star with M$>$4M$_{\odot}$ (Hutchings
et al.\ 1998), similar to the older result for CAL 87 (Cowley et al.\ 1990). 
 
Over the last several years we have obtained photometry of CAL 87,
primarily to confirm the time of minimum light.  The ephemeris has been
very steady over many years, with that given by Schmidtke et al.\ (1993)
being highly accurate.  The time of minimum light is given by: 
\vskip 5pt 
\noindent
\centerline{T$_0=$ HJD 2447506.8021($\pm$0.0002) $+$ 
N$\times$0.4426777($\pm$0.0000016) days}
\vskip 5pt
\noindent
Our observations obtained in 1994 November and 1996 November show that the
depth of the minimum has decreased by over half a magnitude since 1987/88.
The long-term changes in the light curve are discussed in detail by
Hutchings et al.\ (1998). 

\subsection{RX J0925.7$-$4758}

RX J0925 is one of the two galactic supersoft sources we observed.
Although the mean magnitude of the system is relatively bright
($V\sim17.2$), our single spectrum is weak and very noisy below
$\sim$4500\AA\ due to the high interstellar absorption which is estimated
to be more than six magnitudes (Motch 1996).  The emission lines are
relatively weak and narrow compared to other SSS (see Figure 1 and Table
5).  The narrow width of the emission lines and the small observed range
of the optical light curve ($\Delta$m$_V\sim$0.2 mag) suggest RX J0925 is
viewed at a moderately low inclination angle.  However, the fairly large
amplitude of the emission-line velocities, K$_{HeII}=84$ km s$^{-1}$
(Motch, Hasinger, \& Pietsch 1994) means the system must be at a higher
inclination than the nearly face-on systems CAL 83 and RX J0513.  We
estimate the inclination may be in the range $i=30-40^{\circ}$, based on
these factors. 

The distance and hence the luminosity of RX J0925 are poorly known.  Motch
et al.\ estimate E($B-V$)$\sim$2.1 from comparison of the color with RX
J0513.  The strength of the diffuse interstellar bands (marked as ``DIB"
in Figure 1) also suggests E($B-V$)$\sim$2.  Motch (1996) concludes that
the source lies between 1 and 2 kpc, considering the strength of the
interstellar features, the color, and the implied bolometric luminosity. 
This gives RX J0925 an absolute visual magnitude in the range M$_V\sim-1$
to $+1$. 

Motch, Hasinger, \& Pietsch (1994) and Motch (1996) argue from X-ray,
photometric, and spectroscopic data that the orbital period of this system
is near $\sim$3.8 days.  However, alias periods near one day are not yet
completely ruled out, and any one of these would make more sense
physically.  Until a reliable period has been determined from data
spanning several orbital cycles, the parameters for this system must be
considered quite uncertain. 

If the 3.8-day period is used, the data of Motch et al.\ provide only
partial phase coverage.  The fairly long period and large velocity range
together imply the component stars must be widely separated.  For example,
at $i=30^{\circ}$ a 1M$_{\odot}$ compact star would have a 3M$_{\odot}$
companion whose radius must be $\sim8$R$_{\odot}$ in order to fill its
Roche lobe.  Such a star would be substantially evolved since a
main-sequence star of this mass is much too small.  For an inclination of
$i=40^{\circ}$, the secondary mass and required radius are somewhat
smaller but still an evolved star is required.  In either case, the mass
donor would have an absolute magnitude of M$_V\sim0$ and hence should be
visible in the spectrum.  We have looked carefully, particularly in the
long-wavelength region, for absorption lines indicating the presence of an
A--F giant and find none.  This search should be repeated with higher S/N
spectra.  However, if the orbital period is shorter or the velocities not
orbital, the secondary star could be considerably smaller and hence not
visible in the composite spectrum. 

Although the He II and H Balmer emission lines are much weaker than in
most of the other SSS, the strength of the O VI line at 5290\AA\ and the N
III/C III complex at 4630--60\AA\ are comparable.  The strengths of these
lines are likely to be due to the level of ionization in the disk rather
than an abundance effect.  The C/N blend is present in many low-mass X-ray
binaries, even in systems whose location and motion indicate they are Pop
II objects. 

Our spectra show no evidence of displaced lines which would indicate the
presence of outflows or jets.  However, the low signal-to-noise of the
spectrum does not provide a good opportunity to search for these weak
features.  Better spectra need to be obtained for this purpose.

\section{COMPARISON OF THE SSS SPECTRA}

A summary of equivalent widths of selected lines is given in Table 5, as
well as the FWHM of He II 4686\AA.  It is obvious from the intensities of
the H and He II emission lines, particularly those for RX J0513, CAL 83,
and RX J0019, that the Balmer decrement is much stronger than that of the
He II Pickering series so that a substantial fraction of H$\beta$ and the
higher `Balmer' series emission must be attributed to He II.  Although the
height or peak flux of the emission lines varies monotonically from top to
bottom in Table 5 (as shown in Figure 1), the equivalent widths of most
lines are very similar for CAL 83, RX J0019 and CAL 87.  There are
substantial variations in the emission-line widths and in the strength of
the Balmer absorption components.  In particular, the Balmer lines of RX
J0019 and CAL 83 show evidence of strong P-Cygni-type profiles. 

The strength of the C IV doublet at 5801, 5812\AA\ appears to decline
systematically from top to bottom of the sequence shown in Figure 1, but
there is a large variation in the strength of the complex of C III/N III
blend near 4640--50\AA.  The strength of this feature in RX J0925 is
probably not associated with higher CNO abundances in galactic objects
compared to Magellanic Cloud sources since it is very weak in the other
galactic source, RX J0019.  Probably the level of ionization in the inner
disk plays a major role in the strength of this blend.  CHC96 (Fig.\ 7)
show the contribution of various lines in this region to the 4640--50\AA\
complex in RX J0513. In the latter object and CAL 87, the emission extends
to at least 4665\AA, indicating that C III and/or C IV lines are major
contributors.  The strength of the C IV doublet at 5801, 5812\AA\ in RX
J0513 supports this same interpretation.  In contrast, the equivalent
widths of the very high excitation lines of O VI appear to be almost
constant for all systems. 

The emission lines in CAL 87 and SMC 13 are much wider than those of the
other sources, consistent with the known inclinations of CAL 87
($i=78^{\circ}$, eclipsing) and the estimated high inclination of SMC 13
($i\sim75^{\circ}$; Crampton et al.\ 1997).  This implies they are
broadened primarily by Keplerian motions in their accretion disks. 

G\"ansicke et al.\ have discussed the ultraviolet spectra of CAL 83 and RX
J0513, the two most luminous SSS in the LMC.  They find the systems
display very similar spectra, with N V (1239, 1243) and O V (1371) being
the most noticable emission lines in the range 1150--1450\AA.  In both
stars, each component of the N V doublet appears split as if formed in a
rotating ring or disk.  Their small separation supports the other
indications that the inclination in each of these systems in very low. 

\section {HIGHLY SHIFTED LINES AND JETS}

RX J0513 shows widely displaced emission lines on both sides of the
strongest emission features which are thought to arise from bipolar
outflows or jets with velocities of $\sim4000$ km s$^{-1}$ (CHC96,
Southwell et al.\ 1997).  The lower-velocity displaced lines in RX J0019
and the broad 4686\AA\ emission wings in CAL 83 may also indicate high
velocity outflows (Crampton et al.\ 1987, this paper).  In CAL 83 the red
wing of He II 4686\AA, shown in Figure 1, extends to $\sim$2450 km
s$^{-1}$.  The violet components of each line in the CAL 83 appear to have
been absorbed, giving these lines a characteristic P-Cygni profile with an
outflow velocity of $\sim-690$ km s$^{-1}$.  Similar emission and
absorption features are visible in RX J0019 in both the H Balmer and He II
Pickering series, indicating outflows with velocities of $\sim$800 km
s$^{-1}$. 

Livio (1997) has argued by analogy with jets in other astrophysical
objects (from young stellar objects to active galactic nuclei) that highly
collimated outflows or jets occur whenever a central energy/wind source is
surrounded by an accretion disk threaded by a vertical magnetic field. 
Furthermore, he points out that the jet velocities are always the order of
the escape velocity from the central object.  In the case of SSS, the
observed $\sim$700--4000 km s$^{-1}$ velocities are consistent with the
escape velocity from a white dwarf. 

Pringle has suggested that accretion disks which are strongly irradiated
by a central object are unstable to warping, and Southwell, Livio \&
Pringle (1997) discuss the SSS in this context.  As a result of warping,
the associated jets are likely to precess, so that periodic motions of the
shifted lines are expected.  CHC96 showed that their observations of CAL
83 are consistent with such behavior, with a possible period of $\sim$69
days.  Recent data continue to support this. Similarly, in this paper we
have presented evidence that there are long-term changes in the jet
features in RX J0019, although no timescale has yet been established.

\section {DISCUSSION}

The emission-line intensities in the six observed SSS appear to be
strongly correlated with their absolute magnitudes, apparently more so
than with L$_X$ or with the size of the orbit (and hence size of the
accretion disk).  Undoubtedly, the accretion rate plays a major role in the
brightness of the sources, but orientation of the systems must be
significant too.  The high orbital inclinations of CAL 87 and SMC 13 may
account in part for their fainter luminosities. 

The high accretion rate may produce a thick rim extending halfway around
the accretion disk and contributing a substantial fraction of the optical
luminosity (Schandl et al.\ 1996; Meyer-Hofmeister et al.\ 1997). 
Detailed models of SSS light curves by Meyer-Hofmeister et al.\ indicate
that the accretion rate in the RX J0513 system is four times that in RX
J0019 or CAL 87, supporting the idea that the strong emission-line
intensities observed in RX J0513 are a result of high mass transfer.  At
larger accretion rates, the rim of the accretion disk is higher, and
therefore it presents a larger cross section to the compact star.  This
``screen" is illuminated by the compact star, so the contribution from its
integrated brightness is enhanced relative to other contributors to the
total light.  There is increasing evidence that the principal emission
lines arise in a hotspot or ``screen" and hence are directly affected by
the accretion rate.  However, the high excitation O VI lines appear to
show little variation from system to system, or with epoch, and so may
arise from a more stable inner region of the accretion disk.  We note that
the O VI lines weaken or disappear near minimum light in CAL 87 and SMC
13, appearing to be eclipsed.  Unfortunately, the weakness of the O VI
lines and the faintness of the systems preclude accurate velocity
measurements which might reveal the orbital motion of the compact star in
these two systems. 

Long-term spectral variations which seem to be common in all the SSS may
to be due to a combination of irregular variations in the mass-accretion
rate (e.g. RX J0019, CAL 83) and precessing accretion disks (e.g. CAL 83).
According to Meyer-Hofmeister et al.\ the erratic variability in the
long-term light curves can be explained in terms of changing extent and
height of the ``screen".  However, these changes could also result from
variation in the rate of mass transfer.  Such changes could equally well
be responsible for the intensity and profile variations observed in the
emission features.  Simultaneous spectroscopic and photometric monitoring
over several months on systems like CAL 83 and RX J0019 should be helpful
in the study of disk precession. 

A model which successfully reproduces the X-ray and optical luminosities
is that proposed by van den Heuvel et al.\ (1992) in which a 0.7--1.2
M$_{\odot}$ white dwarf accretes matter from a companion star of about
twice its mass at a sufficiently high rate that steady nuclear burning
occurs on its surface.  But, as we have shown, spectroscopic data for many
of these systems indicate that the {\it mass donor has only about half the
mass of the compact star}.  This conflict between the spectroscopic
observations and the currently most popular model needs to be resolved
with improved spectroscopic studies and a fresh look at the models. 

\acknowledgments

APC and PCS acknowledge support from NSF for this work.  We especially
thank the staff of CTIO for their assistance with the observations and 
Mark Wagner for sharing his MMT spectra.

\clearpage

\begin{table}
\caption[]{Comparison of Observed Supersoft Binaries}
\begin{flushleft}
\begin{tabular}{lccccccl}

~~~Name & Period &~~m$_V$  &K$_{HeII}$ & $ ~i$ & ~~M$_V$ & 
L$_{bol}/10^{37}$ 
$^a$  & References \\
&(days) &max(min)  &(km s$^{-1}$) & (deg) &&(erg s$^{-1}$) \\
\hline
RX J0925 & 3.79: & 17.1(17.3) & 84 & 30--40 & ~$\sim0^b$ & 0.03--0.07$^b$ 
& 12,13 \\
CAL 83   & 1.04 & 16.3(17.5) & 35 & 25 & $-$1.3 & 10--100 & 3,5,8,18 \\
RX J0513 & 0.76 & 16.6(17.8) & 11 & 15 & $-$2.0 & 1--60 & 1,6,8,14,15,19 \\
RX J0019 & 0.66 & 12.2(13.0) & 67 & 56 & $+0.6^c$ & 0.3--0.9$^c$ 
& 2,9,10,11,21 \\
CAL 87   & 0.44 & 19.0(20.8) & 73 & 78 & $+$0.3 & 6--20 & 4,16,20 \\
SMC 13   & 0.17 & 20.2(20.6) &100 & 75 & $+$1.4 & 0.8--2  & 7,17 \\     
\hline
\end{tabular}
\end{flushleft}
$^a$ from Greiner (1996) \\
$^b$ for an assumed distance of 1 kpc and E($B-V$)$\sim2.1$ \\
$^c$ for an assumed distance of 2 kpc and E($B-V$)$\sim0.12$ \\
References: 
(1) Alcock et al.\ 1996, 
(2) Beuermann et al.\ 1995, 
(3) Cowley et al.\ 1991, 
(4) Cowley et al.\ 1990, 
(5) Crampton et al.\ 1987, 
(6) Crampton et al.\ 1996, 
(7) Crampton et al.\ 1997, 
(8) G\"ansicke et al.\ 1998,
(9) Greiner \& Wenzel 1995, 
(10) Matsumoto 1996,
(11) Meyer-Hofmeister et al.\ 1997, 
(12) Motch et al.\ 1994, 
(13) Motch 1996, 
(14) Motch \& Pakull 1996
(15) Reinsch et al.\ 1996, 
(16) Schandl et al.\ 1996, 
(17) Schmidtke et al.\ 1996, 
(18) Smale et al.\ 1988, 
(19) Southwell et al.\ 1996, 
(20) van Teesling et al.\ 1996,
(21) Will \& Barwig 1996
	
\end{table}

\clearpage

\begin{table}
\caption[]{1996 Spectroscopic Observations}
\begin{tabular}{lcrc}
~~Name &  ~HJD  &   Exp. &  Phase$^a$  \\
& 2,450,000+  &  (sec) &    (spec) \\
\hline
RX J0513  &  390.733 &   900  &   0.17 \\
          &  390.744 &   900  &   0.18 \\
          &  392.803 &   900  &   0.88 \\
          &  393.788 &   900  &   0.17 \\
CAL 83    &  390.819 &  1000  &   0.66 \\
          &  392.679 &  1200  &   0.44 \\
          &  393.753 &  1200  &   0.47 \\
RX J0019  &  391.496 &    60  &   0.39 \\
          &  393.502 &    60  &   0.43 \\
RX J0925  &  391.859 &  1000  &    - - $^b$ \\
\hline
\end{tabular}
\\ \\
$^a$From maximum positive He II velocity, \\
using ephemerides in text. \\
$^b$Period for RX J0925 too poorly known \\
to calculate phase.\\
\end{table}
 
\clearpage

\begin{table}
\caption[]{Previously Unpublished Photometry of Supersoft X-ray Binaries}
\begin{flushleft}
\begin{tabular}{lcccccc}
\underbar{CAL 83:} \\
Year & HJD & $V$  &$\sigma_V$ & HJD & ~~~~$B$ &$\sigma_B$ \\
& (2400000+) &&& (2400000+) & \\
\hline 
Dec 93 & 49331.558 & 17.472 & 0.010  \\
Nov 95 & 50046.759 & 16.881 & 0.006 \\
Nov 96 & 50389.812 & 17.220 & 0.010 \\
Nov 96 & 50392.742 & 17.342 & 0.008 \\
Nov 96 & 50393.759 & 17.308 & 0.019 & 50393.763 & 17.318 & 0.020 \\
Nov 96 & 50393.829 & 17.282 & 0.014 & 50393.833 & 17.280 & 0.016 \\ 
Nov 96 & 50394.742 & 17.254 & 0.005 & 50394.746 & 17.241 & 0.014 \\
Nov 96 & 50394.837 & 17.248 & 0.014 & 50394.842 & 17.215 & 0.007 \\
\hline \\ \smallskip
\underbar{RX J0925:} \\
Year & HJD & $V$  &$\sigma_V$ & HJD & ~~~~$B$ &$\sigma_B$ \\
& (2400000+) &&& (2400000+) & \\
\hline
Dec 93 & 499330.835 & 17.19 & 0.02  & 499330.843 & 19.15 & 0.02 \\
Dec 93 & 499332.821 & 17.28 & 0.02 & 499332.828 & 19.25 & 0.02 \\
Dec 93 & 499333.843 & 17.31 & 0.04 \\
\hline \\ \smallskip
\underbar{RX J0513:} \\
Year & HJD & $V$  & $\sigma_V$ \\
& (2400000+) &&& (2400000+) & \\
\hline
Nov 95 & 50046.735 & 16.810 &0.007 \\
\hline
\hline
\end{tabular}
\end{flushleft}
	
\end{table}

\clearpage

\begin{table}
\caption[]{Mean Magnitude of CAL 83 at Different Epochs}
\begin{flushleft}
\begin{tabular}{llcl}
Observation UT & mean mag & Ref. & Notes \\
\hline
``1980 \& 1982" & $B=$16.9 to 17.5 & 1 & ($B-V$)=$+0.02$ to $-0.06$ \\
1984 Aug 23-26 & $V=$16.32 & 2  \\
1984 Dec 11-24 & $V=$16.87 & 2 & 0.22 mag orbital amplitude \\
1985 Nov 11-14 & $V=$17.28 & 3 & ($B-V$)$_{mean}=-0.025$ \\
1993 Dec 10 & $V=$17.47 & 4 & only 1 observation \\
1995 Nov 25 & $V=$16.88 & 4 & only 1 observation \\
1996 Nov 2-8 & $V=$17.28 & 4 & ($B-V$)$_{mean}=-0.010$ \\
\hline
\hline
\end{tabular}
\end{flushleft}

References:	
(1) Pakull et al.\ 1985, (2) Smale et al.\ 1988, (3) Crampton \\
et al.\ 1987, (4) this paper

\end{table}

\clearpage

\begin{table}
\caption[]{Emission Line Measurements from 1996 Spectra of 
Supersoft Binaries} 
\begin{flushleft}
\begin{tabular}{lccccccccc}
\multicolumn{1}{l}{~~Name} & 
\multicolumn{8}{c}{Average Equivalent Widths (\AA)} 
 & \multicolumn{1}{l}{FWHM(\AA)}  \\
  \cline{2-9}  
& He II & He II & He II & H$\alpha$ & H$\beta$ & H$\gamma$ & O VI & O VI 
& He II \\
& 4686  & 4541  & 5411  & 6563$^a$ & 4861$^a$ & 4340$^a$ 
& 5290 & 3811$^b$ &  4686  \\
\hline
RX J0513 & $-$22.5 & $-$1.5 & $-$3.1 & $-$41.3 & $-$11.8 & $-$4.1 & $-$0.5 
& $-$1.7 & ~5.8 \\ 
CAL 83   & $-$12.1 & $-$0.8 & $-$1.8 & $-$21.9 & ~$-$4.8 & $-$1.9 & $-$1.3 
& $-$1.6 & ~5.4 \\
RX J0019 & $-$14.1 & $-$1.1 & $-$2.2 & $-$23.6 & ~$-$4.0 & $-$1.1 & $-$0.7 
& $-$0.9 & ~7.4 \\
CAL 87   & $-$14.2 & $-$1.1 & $-$2.4 & $-$19.1 & ~$-$4.3 & $-$1.6 & $-$1.8 
& $-$1.9 & 12.4 \\
RX J0925 & ~$-$5.3  & ~$-$0.5 & $-$0.5 & ~$-$4.1  & ~~.....  & ~..... & $-$0.8 
& ~..... & ~5.1 \\
SMC 13   & ~$-$2.6  & ~$-$0.3 & $-$0.4 & ~$-$6.8  & ~$+$1.3  & $+$1.1 
& $-$0.7 & $-$1.1 & 11.0 \\    
\hline
\hline
\end{tabular}
\end{flushleft}
$^a$ H lines may include a contribution from He II Pickering lines \\
$^b$ weaker O VI line at 3835\AA\ not included in measurement \\
	
\end{table}

\clearpage

\begin{figure}
\caption{Spectra in two wavelength regions: ~a) 3800--5500\AA\ and b)
5100--6600\AA.  The supersoft X-ray binaries, RX J0513.9$-$6951, RX
J0019.8$+$2156, CAL 83, CAL 87, RX J0925.7$-$4758, and SMC 13, are
arranged in order of decreasing emission-line strength from top to bottom.
Some prominent features are identified.  ``DIB'' marks diffuse
interstellar bands.} 
\end{figure}

\begin{figure}
\caption{APT photometry of RX J0019.8$+$2156 from 1995 June to 1996
January, folded on the ephemeris of Will \& Barwig (1996).  The system was
found at three brightness states; each is shown by a different symbol. 
The high-state magnitudes ($\times$) are plotted as observed.  Data from
the medium ($\circ$) and low ($\bullet$) states have been adjusted
brighter by 0.103 and 0.220 mag, respectively.  The shape of the $V$ light
curve does not appear to change from state to state.  The bottom panel
shows the overall ($B-V$)-color curve with no shifts applied since the
mean color does not vary significantly between states.} 
\end{figure}

\begin{figure} 
\caption{The 4600--5000\AA\ region of the spectrum of RX J0513.9$-$6951 in
1996 compared to those observed in 1994 and 1993.  The dashed lines are
located at the positions of the 1994 ``jet'' lines.  These features were
much weaker in 1996.} 
\end{figure}

\begin{figure}
\caption{The spectral regions near H$\alpha$ and H$\beta$ of RX
J0513.9$-$6951 in 1996 compared to the H$\beta$ region in 1994.  The
dashed lines are located at $\pm$4000 km s$^{-1}$ from the main line for
reference purposes.  The extreme right part of the H$\alpha$ plot is
uncertain since it is very close to the end of the observed spectrum. 
Both the violet-displaced absorption and red-displaced emission components
have smaller displacements from the central line at H$\alpha$ compared to
H$\beta$.} 
\end{figure}

\begin{figure}
\caption{Two-dimensional spectrum of CAL 83 in the wavelength region
4600--5050\AA.  The 1\farcs5 slit was oriented E--W centered on CAL 83, and
the resolution along the slit was 0\farcs5 per pixel.  The spatial extent
shown is 93$^{\prime\prime}$, and E is up.  The background light,
including that from the continuum of CAL 83 and from nearby stars, has
been subtracted.  Remnants of some defective CCD columns are visible near
the center.  Note the very different distributions of intensity along the
slit for He II 4686\AA\ compared to H$\beta$ and the [O III] 4959\AA\ and
5007\AA\ lines.} 
\end{figure}

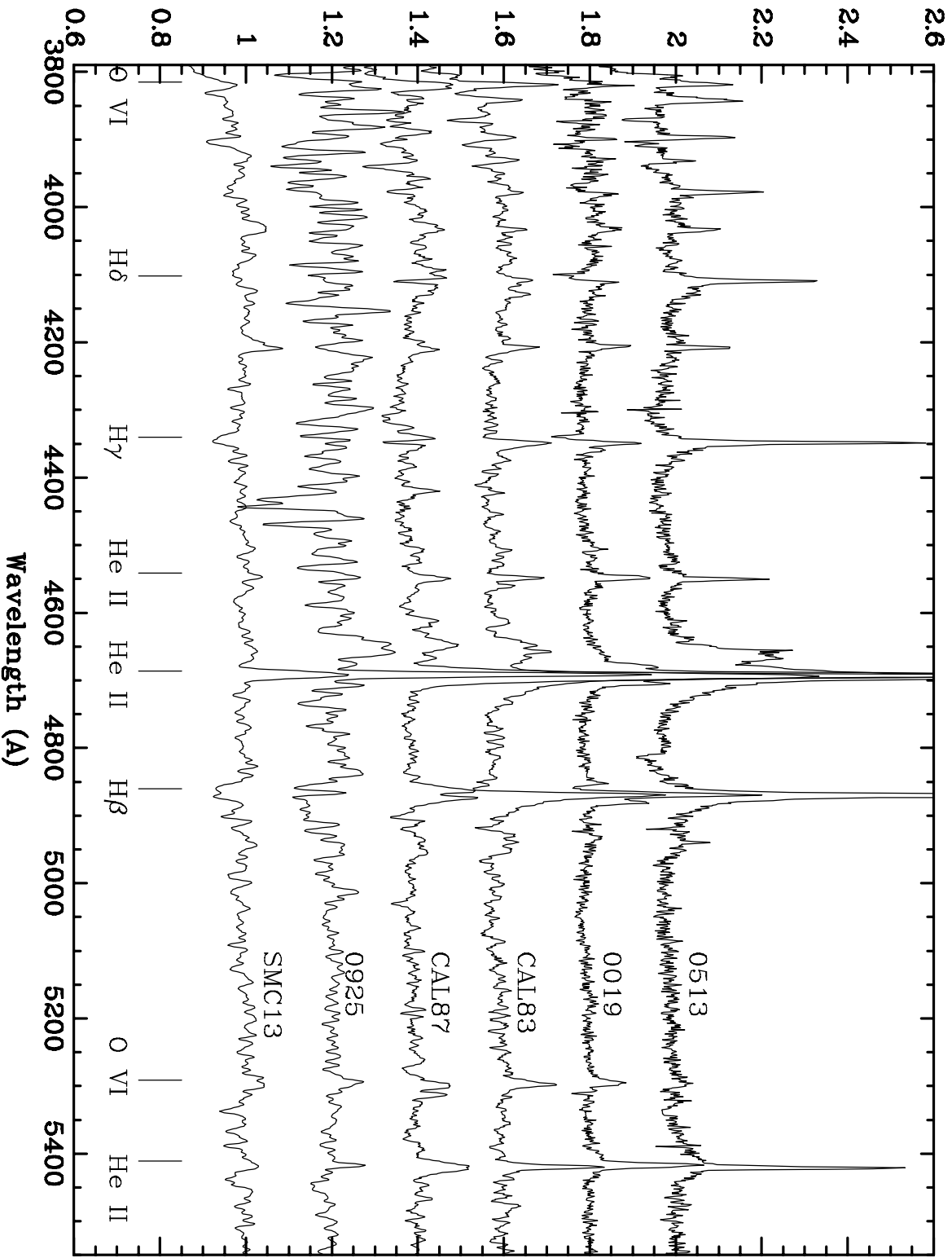
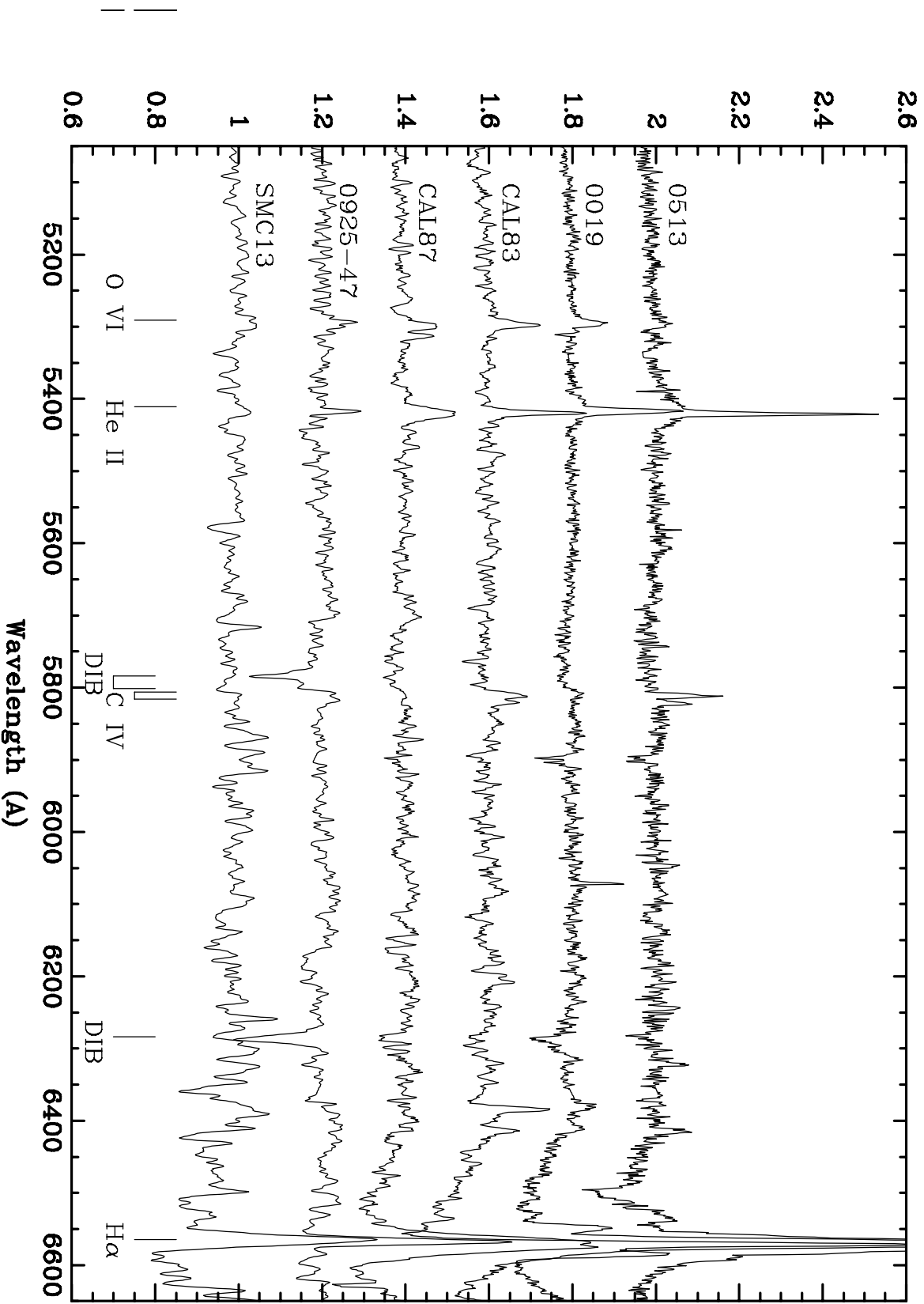
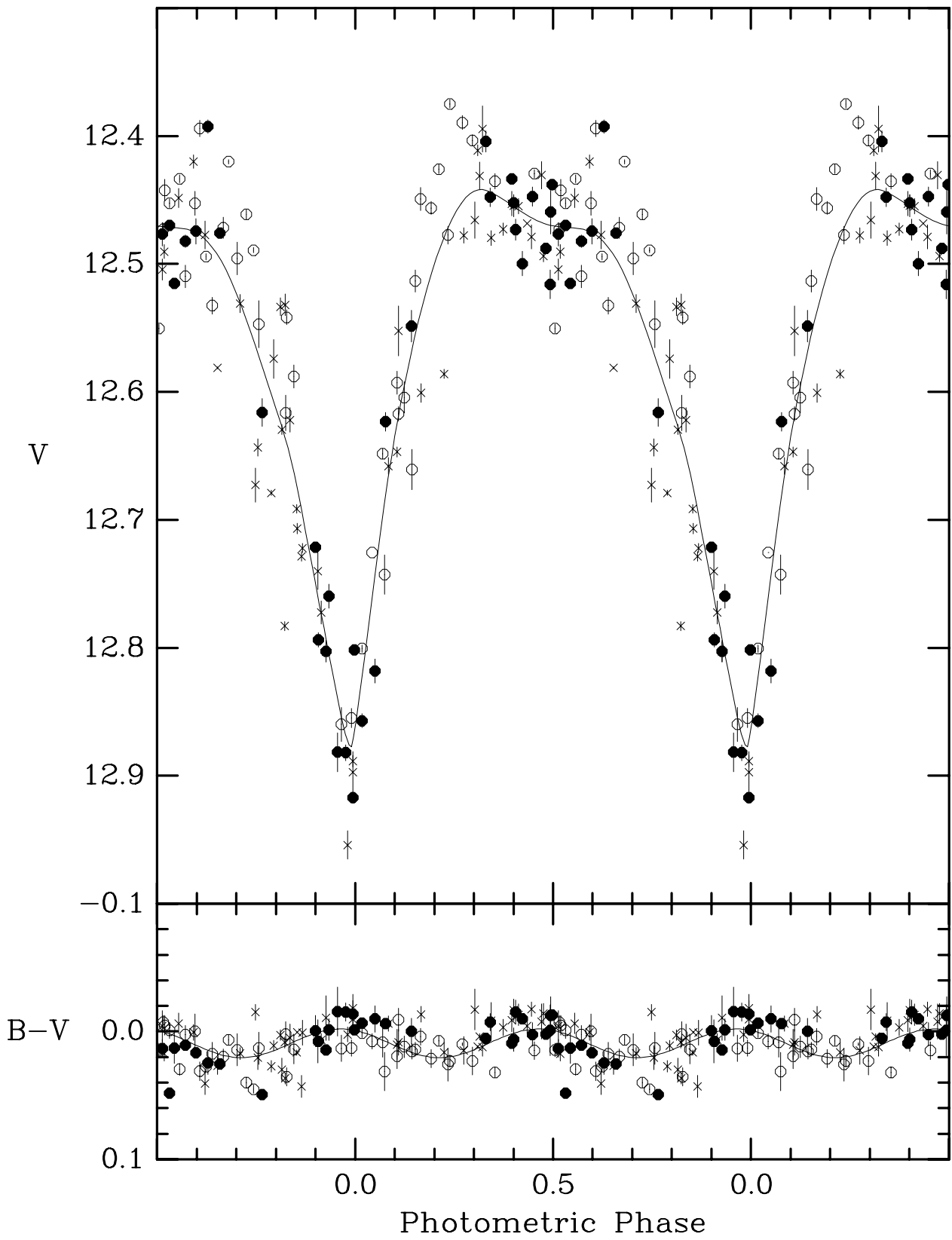
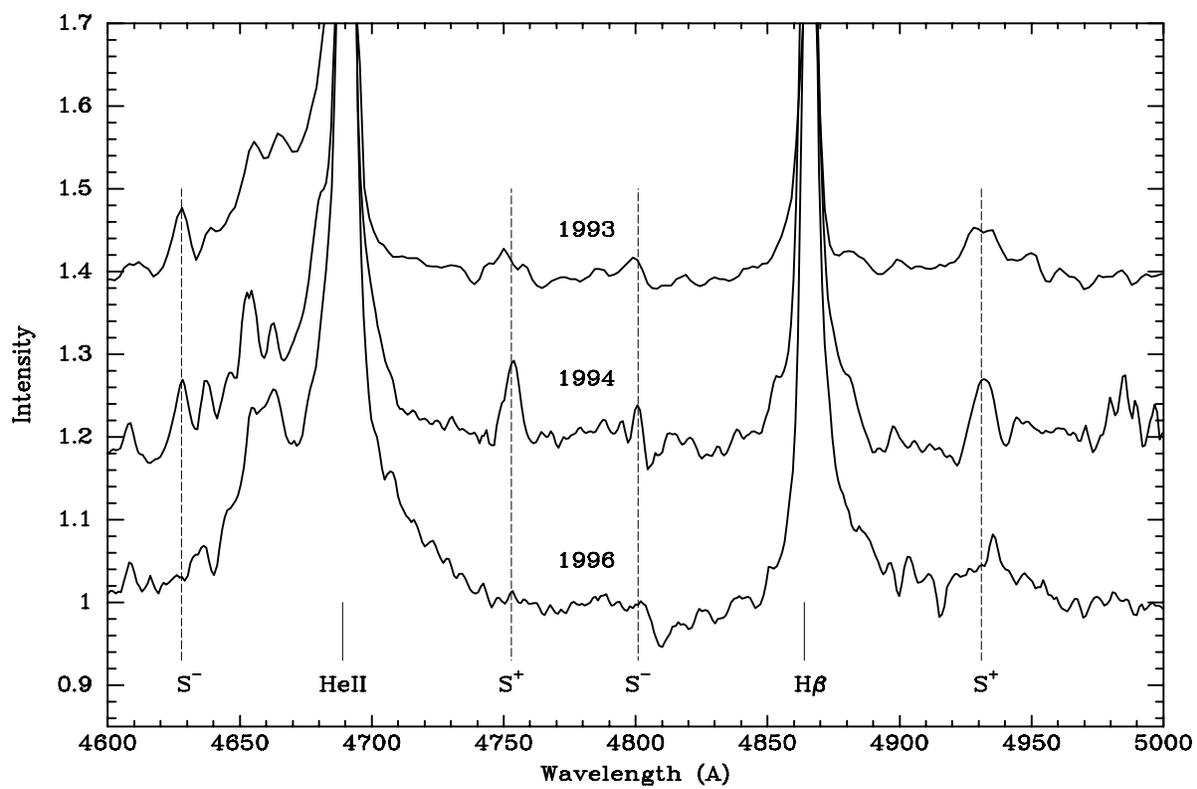
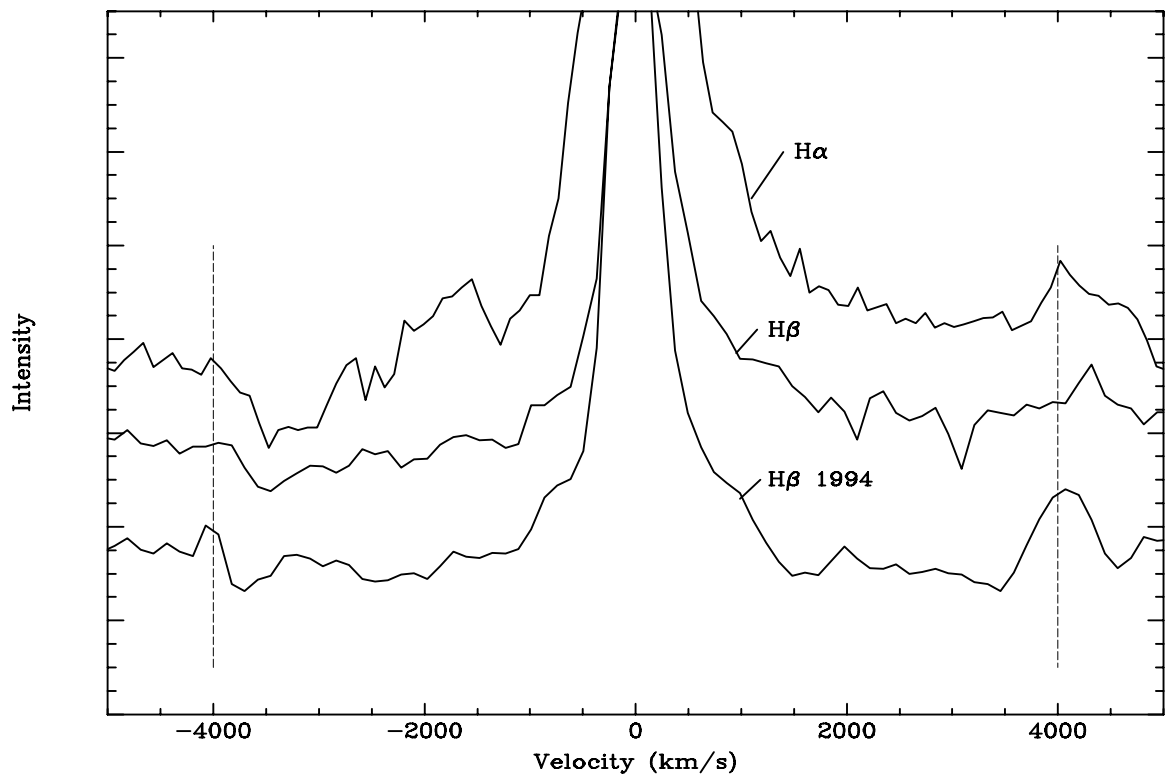
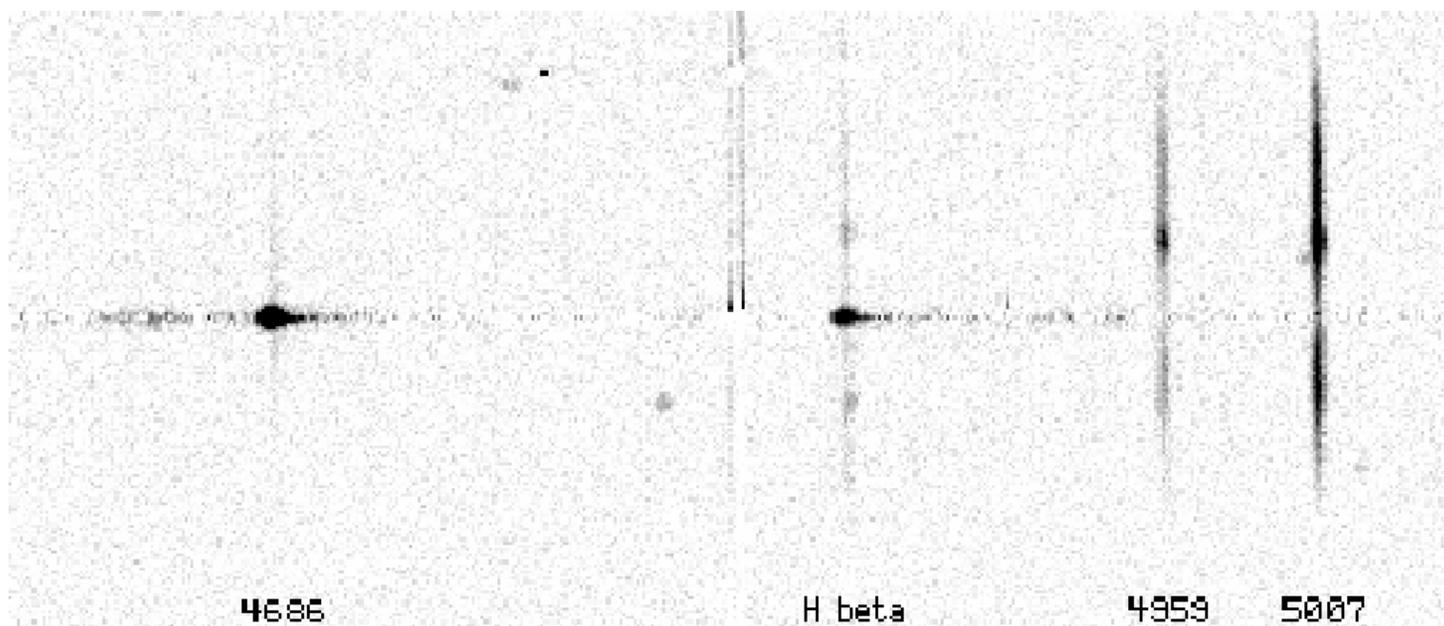
\end{document}